\documentclass[twocolumn,showpacs,preprintnumbers,amsmath,amssymb  amssymb,
 aps,
 prl,
 lengthcheck,
]{revtex4-1}

\usepackage{graphicx}
\usepackage{dcolumn}
\usepackage{bm}
\usepackage{hyperref}
\usepackage{makeidx}
\makeindex
\def\>{\right\rangle}
\def\<{\left\langle}
\def\be{\begin{equation}}
\def\ee{\end{equation}}
\def\ba{\begin{array}{l}}
\def\ea{\end{array}}

\def\beq{\begin{eqnarray}}
\def\eeq{\end{eqnarray}}

\begin{document}

\preprint{APS/123-QED}
\title{Anomalous charge tunneling in the fractional quantum Hall edge states at filling factor $\nu = 5/2$}
\author{M. Carrega$^1$, D. Ferraro$^{2,3,4}$, A. Braggio$^{3}$, N. Magnoli$^{2,4}$, M. Sassetti$^{2,3}$}
 \affiliation{
$^1$ NEST, Istituto Nanoscienze - CNR and Scuola Normale Superiore, I-56126 Pisa, Italy. \\$^2$ Dipartimento di Fisica, Universit\`a di Genova,Via Dodecaneso 33, 16146, Genova, Italy.\\
$^3$ CNR-SPIN, Via Dodecaneso 33, 16146, Genova, Italy.\\ 
$^4$ INFN, Via Dodecaneso 33, 16146, Genova, Italy.}
\date{\today}

\begin{abstract}
  We explain effective charge anomalies recently observed for fractional quantum Hall edge states at $\nu=5/2$ [M. Dolev, Y.
  Gross, Y. C. Chung, M. Heiblum, V. Umansky, and D. Mahalu, Phys.
  Rev. B.  \textbf{81}, 161303(R) (2010)]. The experimental data of
  differential conductance and excess noise are fitted, using the anti-Pfaffian model, by properly take into account renormalizations of the Luttinger parameters induced by the coupling of the system with an intrinsic $1/f$ noise. We demonstrate that a peculiar agglomerate
  excitation with charge $e/2$, double of the expected $e/4$ charge,
  dominates the transport properties at low energies.
  \end{abstract}

\maketitle

\emph{Introduction.}--Since its discovery~\cite{Willett87}, the
fractional quantum Hall (FQH) state at filling factor $\nu=5/2$ has
been subject of intense investigations.  Many proposals have been
introduced in order to explain this exotic even denominator, ranging
from an Abelian description~\cite{Halperin93} to more intriguing ones
which support non-Abelian excitations, like the Moore-Read Pfaffian
model~\cite{Moore91, Fendley07}
or its
particle - hole conjugate, the anti-Pfaffian model~\cite{Lee07}.
The possible applications for the topologically protected quantum computation of non-Abelian excitations 
aroused even more interest for this FQH state~\cite{Nayak08}.\\
In these models the excitations have a fundamental charge $e^*=e/4$ ($e$ the
electron charge). This fact has been experimentally supported by bulk measurements~\cite{Yacoby11} and with current noise 
experiments through a quantum
point contact (QPC) geometry~\cite{Dolev08}, successfully applied for other FQH 
states~\cite{DePicciotto97, 
Chung03}.
Very recently, measurements were reported~\cite{Dolev10} for $\nu=5/2$ where the $e/4$ charge value is observed
at high temperatures, while at low temperatures the measured charge reaches the unexpected value $e/2$. 
Analogous enhancement of the carrier charge has been already observed~\cite{Chung03, Bid09} 
and theoretically explained~\cite{Ferraro08}  
in other composite FQH states belonging 
to the Jain sequence. However, there is still no interpretation of this phenomenon in the $\nu=5/2$ state.\\
In this letter we propose an explanation for these puzzling observations, showing that a different kind 
of excitation, the $2$-agglomerate, with charge double of the fundamental one dominates the transport at low energies. This excitation cannot be simply interpreted in terms of a bunching phenomena of single-quasiparticles due to the non-Abelian nature of the latter.  
We will consider the anti-Pfaffian model,  despite the presented phenomenology could also be consistent with other models. 
In the anti-Pfaffian case three fields are involved, one charged and two neutral (one boson and one Majorana fermion).
The key assumptions of our description are the finite velocity of neutral modes 
and the presence of renormalizations due to the interaction with the external environment. Among all the possible mechanisms leading to a renormalization of the Luttinger parameters \cite{Safi04, Rosenow02} we focus on the effects induced by the 
ubiquitous out of equilibrium $1/f$ noise in presence of a dissipative environment~\cite{DallaTorre10}.

Our predictions show an excellent agreement with experimental data 
on a wide range of temperatures and voltages, demonstrating the validity of the proposed scenario.

\emph{Model.}-- The edge states of $\nu=5/2$ in the anti-Pfaffian model are described as a narrow region at $\nu=3$ 
with nearby a Pfaffian edge of holes with $\nu=1/2$~\cite{Lee07}. Considering the second LL as the "vacuum", the edge is modeled
 as a single $\nu=1$ bosonic branch $\varphi_1$ and a counter-propagating $\nu=1/2$ Pfaffian branch \cite{Fendley07}, composed of a 
 bosonic mode $\varphi_2$ and a Majorana fermion $\psi$. 
 The Lagrangian density is $\mathcal{L}_{\rm edge}=\mathcal{L}_{1}+\mathcal{L}_{2}+
 \mathcal{L}_{\psi}+\mathcal{L}_{12}+\mathcal{L}_{\mathrm{rdm}}$ with ($\hbar=1$) 
\be 
\mathcal{L}_{j}=  \frac{1}{2\pi\nu_j} \partial_{x}\varphi_{j} \left(\eta_j\partial_{t} -v_{j} \partial_{x} \right)\varphi_{j} \qquad j=1,2
\ee 
chiral Luttinger liquid ($\chi$LL) with interaction parameters $\nu_j=1/j$ and velocities $v_j$. The chiralities are $\eta_j=(-1)^{j+1}$ with $\eta=1$  ($\eta=-1$) for a co-propagating (counter-propagating) mode. 
The interaction between the two bosonic modes is 
$\mathcal{L}_{12}= -(v_{12}/2 \pi) \partial_{x} \varphi_{1} \partial_{x} \varphi_{2}$ with $v_{12}$ the coupling strength. 
The term $\mathcal{L}_{\psi}=i \psi (\partial_{t} +v_{\psi} \partial_{x}) \psi$ describes a Majorana fermion propagating with velocity $v_\psi$. 
We also need to include in the Lagrangian a disorder term $\mathcal{L}_{\rm rdm}=\xi(x)\psi e^{i\varphi_1+i2\varphi_2}+{\rm h.c.}$ to describe 
the random electron tunneling 
processes which equilibrate the two branches. The complex tunneling amplitude $\xi(x)$ satisfies  
$\langle \xi(x) \xi^*(x')\rangle=W \delta(x-x')$. These processes bring the edges to equilibrium, recovering the appropriate value of the Hall resistance, in analogy with what happen for $\nu=2/3$  \cite{Kane94,Lee07}.

When the disorder term $\mathcal{L}_{\rm rdm}$ is a relevant perturbation the system is driven 
to a disorder dominated phase \cite{Lee07}. At this fixed point the system naturally decouples in a charged bosonic 
mode with velocity $v_{\mathrm{c}}$ and in two neutral counter-propagating modes (one bosonic and one Majorana fermion) with velocity $v_{\mathrm{n}}$.  Numerical calculations suggest
$v_{\mathrm{n}}<v_{\mathrm{c}}$ \cite{Hu09}.  Related to these
velocities, there are the energy bandwidths
$\omega_{\mathrm{c/n}}=v_{\mathrm{c/n}}/a$, with $a$ a
finite length cut-off. The charged mode bandwidth $\omega_{{\rm c}}$ corresponds to the greatest energy in our model, 
and is assumed to be of the order of the gap. Note that in $\mathcal{L}_{12}$ one has terms that renormalize the neutral and charge mode velocities and terms representing a coupling 
between charge and neutral modes that become irrelevant in this phase \cite{Lee07, Kane94}. 

At the fixed point the Lagrangian density becomes \cite{Lee07}
\beq
\mathcal{L}&=&\frac{1}{2\pi} \partial_{x} 
\varphi_{\rm{c}}(\partial_{t}-v_{\rm{c}} \partial_{x})\varphi_{\rm{c}}+\frac{1}{4 \pi} \partial_{x} \varphi_{\rm{n}} (-\partial_{t} -v_{\rm{n}} \partial_{x})\varphi_{{\rm n}}
\nonumber \\
&&+i\psi \left(\partial_{t} \psi+v_{\rm{n}} \partial_{x}\psi\right)
\label{lagrangian}
\eeq 
with the charged bosonic mode
$\varphi_{\rm{c}}=\varphi_{1}+\varphi_{2}$ related to the electron number density
$\rho(x)=\partial_{x}\varphi_{\rm{c}}(x)/2\pi$ and the neutral
counter-propagating mode $\varphi_{\rm{n}}= \varphi_{1}+2\varphi_{2}$. These bosonic fields satisfy
$\left[\varphi_{\rm{c/n}}(x), \varphi_{\rm{c/n}}(y)\right]=i\pi
\nu_{\rm{c/n}} \mathrm{sgn}(x-y)$ ($\nu_{\rm{c}}=1/2$, $\nu_{\rm{n}}=-1$).

To make the model more realistic, we take into account the effect of the composite nature of the edge interacting with an active substrate and 
the electrical environment. We consider first the 
ubiquitous $1/f$ noise that affects every electrical circuit and that can be generated by trapped charges in the substrate \cite{Paladino02}. If these charges are localized 
near the edge they generate an out of 
equilibrium noise \cite{DallaTorre10} affecting the two bosonic fields $\varphi_1$ and $\varphi_2$ in different  ways. We 
introduce two random sources $f_i$ coupled to the edge densities $\partial_x\varphi_i$, with Lagrangian
$\mathcal{L}_{1/f}=(1/2 \pi) \sum_{i=1,2} f_{i} \partial_{x} \varphi_{i}$ and correlators  
$\langle f_{i}(q,\omega) f^{*}_{i}(q,\omega) \rangle= F_{i}/|\omega|$, with $i=1,2$ \cite{DallaTorre10}.
Dimensional analysis shows that the $1/f$ terms are relevant perturbations, with $F_{i}$ massive parameters.
The external non-equilibrium noise sources heat the system, therefore the stationary condition has to be maintained by the environment through a 
dissipative  cooling mechanism. We model this by means of two baths with 
dissipation rates $\eta_{1}$ and $\eta_{2}$ coupled respectively with $\varphi_{1}$ and $\varphi_{2}$. These dissipative terms called $\mathcal{L}_{\rm bath}$ \cite{Kamenev09} are relevant perturbations with massive coupling constants $\eta_i$.
Generalizing the discussion of  Dalla Torre \emph{et al.} in Ref. \cite{DallaTorre10} to a $\chi$LL case, one can show that, if those terms are sufficiently 
weak $F_{i}, \eta_{i} \rightarrow 0$, in comparison to the other energy scales, but the ratios $F_{i} /\eta_{i}$ remain constant, they become marginal and their effect 
is to modify the Luttinger liquid exponents only. It is worth to note that this result is robust also in presence of counter-propagating modes, and that the considered mechanism doesn't affect the Majorana fermion.

Interestingly the discussed renormalization mechanism is robust against the introduction of disorder that 
doesn't modify the relevance of the massive terms  $\mathcal{L}_{1/f}$ and $\mathcal{L}_{\rm bath}$. Consequently we can consider the effective Lagrangian density on Eq. (\ref{lagrangian}), but with bosonic fields presenting renormalized  $\chi$LL dynamical exponents. Therefore the bosonic Green's function are
 $\langle \varphi_{j}(t) \varphi_{j}(0)\rangle= g_{j} |\nu_{j}| \ln{ (1+i \omega_{j}t)}$  with 
 $g_{j}=g_{j}(F_{1}/\eta_1, F_{2}/\eta_2, F_{1}/F_{2})\geq 1$ 
 ($j=\mathrm{c}, \mathrm{n}$).  A detailed derivation of these facts will be given elsewhere \cite{CarregaNew}. Obviously the renormalizations affect only the 
 dynamical properties of the excitations, without modifying universal 
 quantities like their charge and statistics.\\
\emph{Excitations.}--The generic operator destroying an excitation along
the edge can be written as \cite{Lee07,Nayak08} \be \Psi_{\chi, m,
  n}(x)\propto \chi(x)
e^{i\left[(m/2)\varphi_{c}(x)+(n/2)\varphi_{n}(x)\right]}
\label{Psi}
\ee
here, the integer coefficients $m,n$ and the Ising field $\chi$ define the admissible excitations. In the Ising sector $\chi$ can be $I$ 
(identity operator), $\psi$ (Majorana fermion) or $\sigma$ (spin operator). The operator $\sigma$, due to the non-trivial 
operator product expansion $\sigma \times \sigma=I+\psi$, leads to the non-Abelian statistics of the excitations \cite{Nayak08}. 
The single-valuedness properties of the operators force $m$, $n$ to be even integers for $\chi=I, \psi$ and odd integers 
for $\chi=\sigma$. The charge associated to the operator in Eq. (\ref{Psi}) is $e^{*}_{\chi, m, n}=(m/4)e$ depending on the 
charged mode only. In the following we will indicate an $(m/4)e$ charged excitation as $m$-agglomerate \cite{Ferraro08}. 
The scaling dimension \cite{Kane92} of the operators in Eq. (\ref{Psi}) is
\be
\Delta_{\chi, m, n}=\frac{1}{2}\delta_{\chi}+\frac{g_{\mathrm{c}}}{16} m^{2}+\frac{g_{\mathrm{n}}}{8}n^{2}\,,
\label{Delta}
\ee with $\delta_{I}=0$, $\delta_{\psi}=1$ and $\delta_{\sigma}=1/8$
\cite{Nayak08}.
 Inspection of Eq.
(\ref{Delta}) allows the determination of the more relevant
excitations.  Among all the single-quasiparticle (qp), with charge
$e^{*}=e/4$, the most dominant are $\Psi^{(1)}=\Psi_{\sigma, 1,\pm1}$
with scaling dimensions $\Delta^{(1)}=\Delta_{\sigma, 1,\pm 1}=(g_{\rm
  c}+2 g_{\rm n}+1)/16$.  The other most relevant excitation is the
$2$-agglomerate with charge $2e^*=e/2$ and operator
$\Psi^{(2)}=\Psi_{I,2,0}$ with scaling dimension
$\Delta^{(2)}=\Delta_{I, 2,0}=g_c/4$. It is worth to note that also
the operator $\Psi_{\psi,2,0}$ has a charge $e/2$, but is less
relevant because its scaling dimension is increased by the Majorana
fermion contribution.  All other excitations are less relevant and
will be neglected in the following. 

In the unrenormalized case $(g_{\rm{c}}=g_{\rm{n}}=1)$ the single-qp $(\Psi^{(1)})$ and the
$2$-agglomerate $(\Psi^{(2)})$ have the same scaling dimension, equal
to $1/4$. Renormalization effects qualitatively change the above
scenario. In particular, for $g_{\mathrm{\rm c}}<(1+2g_{\mathrm{\rm
    n}})/3$, the $2$-agglomerate becomes the most relevant excitation
at low energies opening the possibility of a crossover between the two excitations, in agreement with experimental observations. 
Note that, due to the peculiar fusion rules of the $\sigma$ operator, the
$2$-agglomerate cannot be simply created combining two single-qp
without introducing also an excitation with a Majorana fermion in the
Ising sector. This fact suggests that, in the non-Abelian
models, the $2$-agglomerate is not simply given by a bunching of two quasiparticles, namely in general $\Psi^{(2)}\neq (\Psi^{(1)})^{2}$.\\
\emph{Transport properties.}--In the QPC geometry tunneling of
excitations between the two side of the Hall bar is allowed, and can
be described through the Hamiltonian
$H_{\rm{T}}=\sum_{m=1,2} t_{m}{{\Psi}^{(m)}_{R}}^{\dagger}(0)\Psi^{(m)}_{L}(0)+\mathrm{h.c.}
$ where $R$ and $L$
indicate respectively the right and the left edge, $t_{m}$
$(m=1,2)$ the tunneling amplitudes.  Without loss of generality,
we assume the tunneling occurring at $x=0$. At lowest order in
$H_{\rm{T}}$ \cite{Bena06} the backscattering current is $I_{B}=\sum_{m=1,2}
\langle I^{(m)}_{B}\rangle$ with 
\be \langle I^{(m)}_{B}\rangle=m e^{*}\left(1- e^{-\frac{me^{*}
      V}{k_{B} T}}\right)\Gamma_{m}(m e^{*}V)\label{current} 
\ee 
being $V$ the bias,
$T$ the temperature and where
$\Gamma_{m}(E)$ indicates the first order Fermi's Golden rule tunneling rate.
The differential backscattering conductance is given
by $G_{B}=\sum_{m=1,2} G^{(m)}_{B}$
with $G^{(m)}_{B}=d\langle I^{(m)}_{B}\rangle/dV$.\\
Current noise~\cite{Chamon96,Bena06} is another relevant
quantity in order to provide information on the $m$-agglomerate
excitations. The finite frequency symmetrized 
noise is $S_{B}(\omega)=\int_{-\infty}^{+\infty}dt e^{-i\omega t}\langle \{\delta I_B(t),
\delta I_B(0)\}\rangle$ with 
$\delta I_B=I_B-\langle I_B \rangle$ with $\{\cdot,\cdot\}$ the anticommutator. At lowest order in the tunneling, 
is simply given
by the sum of the two contributions $S_{B}(\omega)=\sum_{m=1,2}
S^{(m)}_{B}(\omega)$ with 
\be S^{(m)}_{B}(\omega)=(m
e^{*})\sum_{\epsilon=\pm} \coth{\left[\frac{\epsilon\omega
      +m\omega_{0}}{2k_{B}T}\right]} I_{B}(\epsilon\omega+
m\omega_{0})
\label{FFNoise}
\ee where $\omega_{0}=e^{*}V$. 
A detailed analysis of this quantity will be given elsewhere \cite{CarregaNew}, in this letter we will focus only on the zero frequency limit. One can introduce
the backscattering current excess noise $S_{\mathrm{exc}}=S_{B}(0)-4 k_{B}T G_{B}|_{V=0}$ that, in the lowest order in the tunneling, 
can be directly 
compared with the current noise measured in the experiments.\\
\emph{Results}--We will compare now the theoretical predictions
with the raw experimental data for the differential
conductance and the excess noise in the extreme weak-backscattering
regime, taken from Ref.~\onlinecite{Dolev10}. In the shot noise regime $k_{B}T\ll e^{*}V$ the current in Eq. (\ref{current}) follows
specific power-laws $I_{\mathrm{B}}\propto V^{\alpha-1}$. Being in the shot regime one has the same power-laws in the
excess noise $S_{\mathrm{exc}}\propto V^{\alpha-1}$.
The exponent $\alpha$ changes varying the voltages and it is related to the scaling dimensions in Eq. (\ref{Delta}). 
In particular it is $\alpha=g_{\mathrm{c}}$ at very low energy, where the $2$-agglomerate 
dominates. At higher voltages, where the single-qp dominates, it 
is possible to distinguish two different regimes. For $e^{*}V \ll \omega_{\mathrm{n}}$, 
where the neutral modes contribute to the dynamics,  one 
has $ \alpha=g_{\mathrm{c}}/4+g_{\mathrm{n}}/2+1/4$, while for $e^{*}V \gg \omega_{\mathrm{n}}$ 
the neutral modes are uneffective and the exponent reduces to $\alpha=g_{\mathrm{c}}/4$. 
In thermal regime $k_{B}T\gg e^{*}V$ the conductance is independent on
the voltage and scales with temperature like $G_{\mathrm{B}}|_{V=0}\propto
T^{\alpha-2}$ while $S_{\mathrm{exc}}\propto V^{2}$.
\begin{figure}
\centering
\includegraphics[width=0.38\textwidth]{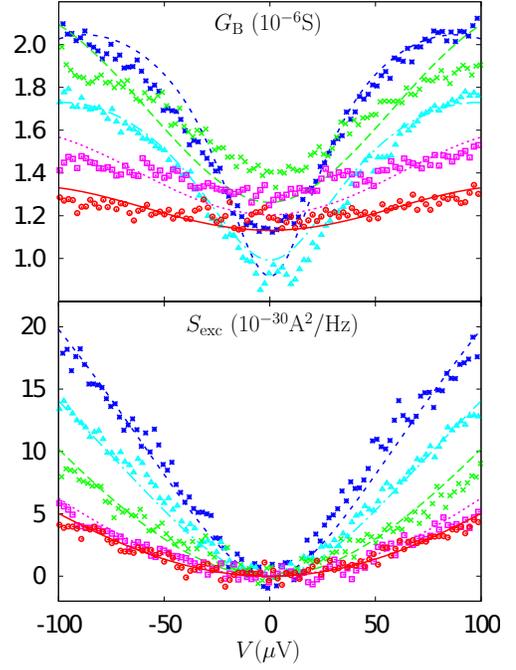}
\caption{Differential conductance $G_{\mathrm{B}}$ (top) and excess
  noise $S_{\mathrm{exc}}$ (bottom) as a function of voltage. Symbols
  represent the experimental data, corresponding to the sample
  indicated with the full circles in Fig. 5 of Ref. \cite{Dolev10},
  with courtesy of M. Dolev. Different styles indicate different
  temperatures: $T=27$ mK (asterisks, short-dashed blue), $T=41$ mK
  (triangles, dashed-dotted cyan), $T=57$ mK (crosses, long-dashed
  green), $T=76$ mK (squares, dotted magenta), $T=86$ mK (circles,
  solid red). Fitting parameters are: $g_{\mathrm{c}}=2.8$,
  $g_{\mathrm{n}}=8.5$, $\omega_{\mathrm{c}}=500$ mK,
  $\omega_{\mathrm{n}}=150$ mK ($k_{B}=1$). $\gamma_{1}=|t_{1}|^{2}/(2 \pi v_{\mathrm{c}})^{2}=3.1\cdot
  10^{-2}$, $3.3\cdot 10^{-2}$, $5.6\cdot 10^{-2}$, $4.9\cdot
  10^{-2}$, $4.2\cdot 10^{-2}$ and $\gamma_{2}=|t_{2}|^{2}/(2 \pi v_{\mathrm{c}})^{2}=1.2\cdot 10^{-2}$,
  $7.6\cdot 10^{-3}$, $1.7\cdot 10^{-3}$, $4.9\cdot 10^{-5}$,
  $4.2\cdot 10^{-5}$.}
\label{fig:FigA}
\end{figure}
In Fig. \ref{fig:FigA} we show experimental data and theoretical
predictions for the backscattering differential conductance (top) and
excess noise (bottom) at different temperatures. All curves are
obtained fitting with the same values for the
renormalization parameters $(g_{\rm c}= 2.8,g_{\rm n}= 8.5)$ and neutral mode bandwidth ($\omega_{{\rm n}} =150$ mK). We also assume that the
tunneling coefficients associated to the single-qp ($\gamma_{1}$) and the 
$2$-agglomerate ($\gamma_{2}$) could vary with temperature. 
The fitting has been validated by means of the standard $\chi^{2}$ test 
and shows an optimal agreement with the whole sets of data. Notice that the value of the 
neutral mode bandwidth is lower than $\omega_{{\rm c}} = 500$ mK, 
which is of the order of the gap, according with the Ref. \cite{Hu09}.\\
The backscattering differential conductance always presents a minimum at zero 
bias which is the signature of the 'mound-like' behavior generally observed for 
the transmission in the QPC geometry at very weak-backscattering \cite{Dolev08}.  
For low enough temperatures, i.e. blue (short dashed) and cyan (long dashed) lines, one can see the 
dominance of the $2$-agglomerate for low bias $V\lesssim50$ $\mu$V and a 
crossover region related to the dominance of the single-qp increasing voltages.  
At higher temperatures, where the single-qp contribution becomes relevant, the curves appear quite flat and voltage independent (dotted magenta 
and solid red lines). This is a signature of the ohmic behaviour reached in the thermal regime $e^{*}V \ll k_{B}T$. Notice that the 
presence of renormalizations for the charged and neutral modes is crucial in the fit.\\
Let us discuss now the excess noise curves. At high temperature (low bias) they
present an almost parabolic behavior as expected for the thermal
regime.
Nevertheless this behavior is also present for $e^*V\gg k_B T$. This effect is not universal and it is due to the peculiar scaling dimension of the $2$-
agglomerate and to the value of the charge mode renormalization. At high bias ($V\approx 100$ $\mu$V) the lowest temperature curve deviates 
from the quadratic behavior as a consequence of the single-qp contribution. \\
In Fig. \ref{fig:FigB} we compare the effective charge
$e_{\mathrm{eff}}$ (triangles), calculated from our theoretical curves
using a single parameter fitting procedure, with the results of Ref. \cite{Dolev10} (circles with error bars). This result reinforces the idea that the evolution of the effective charge, as a function of the temperature, is essentially due to the crossover between the single-qp and the $2$-agglomerate contributions.\\
\begin{figure}
\centering 
\includegraphics[width=0.38\textwidth]{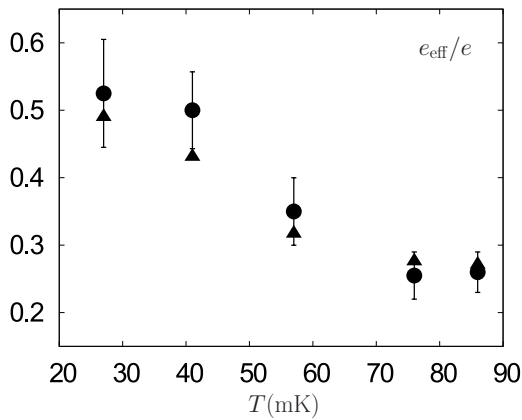}
\caption{Effective charge, in unit of the electron charge, as a function of temperature. 
Circles with error bars are the experimental data of Fig. 5 of Ref. \cite{Dolev10}, with courtesy of M. Heiblum. Triangles are the effective charges 
obtained from the theoretical curves of excess noise in Fig.\ref{fig:FigA}.}
\label{fig:FigB}
\end{figure}
\emph{Conclusions.}-- We fit recent experimental data on
differential backscattering conductance and excess noise in a quantum
point contact geometry for filling factor $\nu=5/2$ in the weak back-scattering regime, demonstrating that the tunneling excitation has a charge double of the fundamental one at very low 
temperatures. 
In order to fit the experimental data, it is essential to assume the presence of interactions which renormalize
the scaling behavior. We present a model for them in terms of the coupling of the system with the ubiquitous 
non-equilibrium $1/f$ noise. This external coupling only affects the dynamical properties of the system, such as the scaling dimension, but does not 
change universal quantities, i.e. charge and statistics of the excitations.
The presented phenomenology is also consistent with other models for the $\nu=5/2$ state and it is not restricted to the considered anti-Pfaffian model. 

\emph{Acknowledgements.}- We thank Y. Gefen, Y. Oreg, A. Cappelli and
G. Viola for valuable discussions. A particular acknowledge is for the
experimental group of M. Heiblum for providing us the raw data of
their experiment and for the kind hospitality of one of us (A.B.). We acknowledge the support of the CNR STM 2010 programme and the EU-FP7 via ITN-2008-234970 NANOCTM.


\begin{thebibliography}{10}
\bibitem{Willett87} R. Willett, J. P. Eisenstein, H. L. Stormer, D. C. Tsui, A. C. Gossard, and J. H. English, Phys. Rev. Lett. \textbf{59}, 1776, (1987).
\bibitem{Halperin93} B. I. Halperin, P. A. Lee, and N. Read, Phys. Rev. B \textbf{47}, 7312 (1993); 
\bibitem{Moore91}G. Moore and N. Read, Nucl. Phys. B \textbf{360}, 362 (1991);  R. H. Morf, Phys. Rev. Lett. \textbf{80}, 1505 (1998).
\bibitem{Fendley07} P. Fendley, M. P. A. Fisher, and C. Nayak, Phys. Rev. B \textbf{75}, 045317 (2007).
\bibitem{Lee07} S. -S. Lee, S. Ryu, C. Nayak, and M. P. A. Fisher, Phys. Rev. Lett. \textbf{99}, 236807 (2007); M. Levin, B. I. Halperin, and B. Rosenow, 
Phys. Rev. Lett. \textbf{99}, 236806 (2007).
\bibitem{Nayak08} C. Nayak, S. H. Simon, A. Stern, M. Freedman, and  S. Das Sarma, Rev. Mod. Phys. \textbf{80}, 1083 (2008).
\bibitem{Yacoby11} V. Venkatachalam, A. Yacoby, L. Pfeiffer, and K. West, Nature \textbf{469}, 185 (2011).
\bibitem{Dolev08} M. Dolev, M. Heiblum, V. Umansky, A. Stern, and D. Mahalu,  Nature \textbf{452}, 829 (2008).
\bibitem{DePicciotto97} R. de Picciotto, M. Reznikov, M. Heiblum, V. Umansky, G. Bunin, and D. Mahalu, Nature \textbf{389}, 162 (1997); L. 
Saminadayar, D. C. Glattli, Y. Jin, and B. Etienne, Phys. Rev. Lett. \textbf{79}, 2526 (1997).
\bibitem{Chung03} Y. C. Chung, M. Heiblum, and V. Umansky, Phys. Rev. Lett. \textbf{91}, 216804 (2003).
\bibitem{Dolev10} M. Dolev, Y. Gross, Y. C. Chung, M. Heiblum, V. Umansky, and D. Mahalu, Phys. Rev. B. \textbf{81}, 161303(R) (2010).
\bibitem{Bid09} A. Bid, N. Ofek, M. Heiblum, V. Umansky, and D. Mahalu, Phys. Rev. Lett. \textbf{103}, 236802 (2009).
\bibitem{Ferraro08} D. Ferraro, A. Braggio, M. Merlo, N. Magnoli, and M. Sassetti, Phys. Rev. Lett. \textbf{101}, 166805 (2008); D. Ferraro, A. Braggio, 
N. Magnoli, and M. Sassetti, New J. Phys. \textbf{12}, 010312 (2010); D. Ferraro, A. Braggio, N. Magnoli, and M. Sassetti, Phys. Rev. B \textbf{82}, 
085323 (2010).
\bibitem{Safi04} M. Sassetti and U. Weiss, Europhys. Lett. \textbf{27}, 311 (1994); I. Safi and H. Saleur, Phys. Rev. Lett. \textbf{93}, 126602, (2004); A. H. Castro Neto, C. de C. Chamon, and C. Nayak, Phys. Rev. Lett. 
\textbf{79}, 4629, (1997).
\bibitem{Rosenow02} B. Rosenow and B. I. Halperin, Phys. Rev. Lett. \textbf{88}, 096404 (2002); E. Papa and A. H. MacDonald, Phys. Rev. Lett. \textbf{93}, 126801 (2004); K. Yang, Phys. Rev. Lett. \textbf{91}, 036802 (2003).
\bibitem{DallaTorre10} E. G.  Dalla Torre, E. Demler, T. Giamarchi, and  E. Altman, Nature Physics \textbf{6}, 806 (2010).
\bibitem{Kane94} C. L. Kane, Matthew P. A. Fisher, and J. Polchinski, Phys. Rev. Lett. \textbf{72}, 4129 (1994).
\bibitem{Hu09} Z.-X. Hu, E. H. Rezayi, X. Wan, and K. Yang, Phys. Rev. B \textbf{80}, 235330 (2009).
\bibitem{Paladino02} J. Muller, S.  von Molnar, Y. Ohno, and H. Ohno, Phys. Rev. Lett. \textbf{96}, 186601 (2006);  E. Paladino, L. Faoro, G. Falci, and R. Fazio, Phys. Rev. Lett. \textbf{88}, 228304 (2002).
\bibitem{Kamenev09} A. Kamenev, A. Levchenko, Adv. Phys. \textbf{58}, 197 (2009).
\bibitem{CarregaNew} M. Carrega, D. Ferraro, A. Braggio, N. Magnoli, and M. Sassetti \emph{in preparation}.
\bibitem{Kane92} C. L. Kane and M. P. A. Fisher, Phys. Rev. Lett. \textbf{68}, 1220 (1992).
\bibitem{Bena06} C. Bena and C. Nayak, Phys. Rev. B \textbf{73}, 155335 (2006)
\bibitem{Chamon96} C. de C. Chamon, D. E. Freed, and X. G. Wen, Phys. Rev. B \textbf{53}, 4033 (1996); C. Bena and I. Safi, Phys. Rev. B \textbf{76}, 125317 (2007).
\end{thebibliography}
\end{document}